\documentclass[twocolumn,amsfonts,amsmath,amssymb,final,aps,prd,footinbib,longbibliography,nofootinbib,10pt]{revtex4-2}
\usepackage[dvipsnames]{xcolor}
\usepackage{dcolumn}
\usepackage[colorlinks=true,allcolors=Blue]{hyperref}
\usepackage{upgreek,graphicx,manfnt,pifont,colonequals,bbm,stmaryrd}
\usepackage[shortlabels]{enumitem}

\def\be{\begin{eqnarray}}
\def\ee{\end{eqnarray}}
\def\bea{\begin{eqnarray}}
\def\eea{\end{eqnarray}}

\newcommand{\mtbE}{\mathbb E}

\begin{document}

\title{Schur Indices of Class \texorpdfstring{$\mathcal{S}$}{S} and Quasimodular Forms}

\medskip

\author{Christopher Beem}
\affiliation{Mathematical Institute, University of Oxford, Woodstock Road, Oxford, OX2 6GG, United Kingdom}

\author{Palash Singh}
\affiliation{Mathematical Institute, University of Oxford, Woodstock Road, Oxford, OX2 6GG, United Kingdom}

\author{Shlomo S. Razamat}
\affiliation{Department of Physics, Technion, Haifa 32000, Israel}



\begin{abstract}
\noindent We investigate the unflavoured Schur indices of class $\mathcal S$ theories of modest rank, and in the case of $\mathcal{N}=4$ super Yang--Mills theory with special unitary gauge group of somewhat more general rank, with an eye towards their modular properties. We find closed form expressions for many of these theories in terms of quasimodular forms of level one or two, with the curious feature that in general they are sums of quasimodular forms of different weights. For type $\mathfrak{a}_1$ theories, the index can be fixed by taking a simple Ansatz for the family of quasimodular forms appearing in the expansion of this type and demanding that the result be sufficiently regular as $q\to0$. For higher rank cases, an equally simple construction is lacking, but we nevertheless find that these indices can be expressed in terms of mixed-weight quasimodular forms.
\end{abstract}

\maketitle


The Schur limit of the superconformal index of a four-dimensional $\mathcal{N}=2$ superconformal field theory \cite{Gadde:2011uv,Gadde:2011ik} is, in general, the vacuum character of a quasi-Lisse vertex operator algebra (VOA) \cite{Beem:2013sza}, and as such is the solution of a weight zero modular linear differential equation (MLDE) \cite{Beem:2017ooy,Arakawa:2016hkg}. As a result, it is in general a component of a (potentially logarithmic) vector-valued modular function. Though there are now a good number of examples where these characters are understood concretely at the level of modular functions (such as when the associated VOA is rational, or an admissible level affine Kac--Moody VOA), the appropriate characterisation of the modular objects arising in the general case remains unclear.

A potentially illustrative set of examples where exact expressions for unflavoured Schur indices are known are the rank-one $F$-theory SCFTs. For these theories, the MLDE for the Schur index is of order two and can be solved exactly \cite{Beem:2017ooy,Arakawa:2016hkg}, for original works see \cite{KanekoZ:1998,KanekoK:2003,KanekoNS:2013}. For the $\mathfrak{a}_1$ and $\mathfrak{a}_2$ theories, which are Argyres--Douglas theories, the corresponding indices can be written as a power of the Dedekind $\eta$ function\footnote{We define various relevant special functions and key constructions involving modular forms more generally in the appendix.} times holomorphic modular forms for the congruence subgroups $\Gamma_0^0(3)$ and $\Gamma(2)$, respectively,
\begin{equation}\label{eq:a1a2_AD_indices}
    \begin{aligned}
            \mathcal{I}^{{\rm rank}\, 1}_{\mathfrak{a}_1}(q)&=\left(\frac{\eta(3\tau)}{\eta(\tau)}\right)^3~,\\
            \mathcal{I}^{{\rm rank}\, 1}_{\mathfrak{a}_2}(q)&=\left(\frac{\eta(2\tau)}{\eta(\tau)}\right)^8~.
    \end{aligned}
\end{equation}
On the other hand, for $\mathfrak{d}_4$ and $\mathfrak{e}_{6,7,8}$ theories, all of which are realised in class $\mathcal{S}$ with regular punctures \cite{Gaiotto:2009we}, the indices are (again, up to overall powers of the $\eta$ function) \emph{quasi}modular forms for the full modular group $\Gamma_1$, \emph{i.e.}, polynomials in the Eisenstein series $\mtbE_{2,4,6}(\tau)$. For example, one has \cite{KanekoNS:2013,Arakawa:2016hkg}
\begin{equation}\label{eq:d4e6_rank_one_indices}
    \begin{aligned}
        \mathcal{I}^{{\rm rank}\,1}_{\mathfrak{d}_4}(q) &=\frac{3\;\mtbE_4'(\tau)}{\eta(\tau)^{10}}=\frac{42\mtbE_6(\tau)-12\mtbE_2(\tau)\mtbE_4(\tau)}{\eta(\tau)^{10}}~,\\
        \mathcal{I}^{{\rm rank}\,1}_{\mathfrak{e}_6}(q) &=\frac{\Delta(\tau)+90720\;\mtbE_6(\tau)\mtbE_4'(\tau)}{11\;\eta(\tau)^{22}}~,
    \end{aligned}
\end{equation}
and the expressions for the $\mathfrak{e}_7$ and $\mathfrak{e}_8$ theories are more complicated (but still quasimodular).

Including the $\eta$ factors, both of these examples would rightly be described as quasimodular of weight \emph{one} and depth one. However, from the perspective of the associated VOA, the most natural modular transformation to consider for these indices is the weight zero transformation. Indeed, as we review later, the weight $k$ modular action on a quasimodular form of weight $k$ and depth $p$ generates a polynomial of degree $p$ in $c/(c\tau+d)$ with quasimodular coefficients, so with respect to the weight $k-p$ action one has a logarithmic vector-valued modular form (with $(\log q)$'s in numerators rather than in denominators). 

There is a notable difference between the way that the indices in \eqref{eq:a1a2_AD_indices} and those in \eqref{eq:d4e6_rank_one_indices} realise vector-valued modular objects. In the former, it is through holomorphic modular forms for congruence subgroups that are therefore vector valued when viewed with respect to the full modular group. Alternatively, for the latter (which are logarithmic) it is through quasimodular forms for the full modular group, namely Eisenstein series including $\mtbE_2(\tau)$, which give rise to vector-valued functions due (morally) to the anomalous transformation of the second Eisenstein series.

In going on to consider more general SCFTs whose Schur indices satisfy much higher order MLDEs, it would be natural to expect a combination of these mechanisms---quasimodularity and modularity for congruence subgroups of large index---to be at play. In this paper, we find compelling evidence that for fairly general class $\mathcal{S}$ Schur indices this is not the case, and rather it is only a mechanism more akin to the quasimodularity of the above examples giving rise to the vector valued modular functions in question. More precisely, we have found that the Schur indices of large families of class $\mathcal{S}$ theories can be written as $\eta$ function prefactors multiplying holomorphic quasimodular forms for the full modular group \footnote{Here, full modular group means the modular group under which the associated MLDE would be invariant. So for Schur indices which are $q$-series this is the full modular group ($\Gamma_1$), while for those with half-integer powers of $q$ this is the congruence subgroup $\Gamma^0(2)$ \cite{Beem:2017ooy}.}. Strikingly, in the general case the holomorphic modular form in question is of \emph{mixed weight}. From these expressions it is simple to deduce the full modular transformations of these indices as well as the dimension of their corresponding modular vector, \emph{i.e.}, the order of the LMDE that should be solved by the index.

In the next section, we present many exact expressions for class $\mathcal{S}$ and $\mathcal{N}=4$ Schur indices in terms of holomorphic quasimodular forms. In the case of $\mathfrak{a}_1$ theories, we propose a simple (conjectural) procedure to fix the exact index in any example in terms of quasimodular forms by positing an Ansatz whose coefficients are fixed by a straightforward computation. In the following section, we comment on the general structural features of these indices that can be inferred from these examples. We propose an Ansatz generalising that of the $\mathfrak{a}_1$ case for $\mathfrak{a}_2$ theories with arbitrary numbers of maximal and minimal punctures, albeit one whose free coefficients are less easily fixed. We offer some thoughts about future directions in a summary section. Some relevant technical details regarding modular and quasimodular forms are included in the sole appendix.

\medskip

\noindent \emph{Authors' note: while this work was being completed, we learned of the work \cite{PanPeelaers}, which has significant overlap with our empirical results and indeed derives expressions for many Lagrangian class $\cal S$ Schur---in particular those of type $\mathfrak{a}_1$---by direct evaluation of the relevant contour integrals.}

\section{\label{sec:exact_schur_indices}Exact Schur Indices}

In this first section, we collect and discuss a number of cases where the unflavoured Schur index can be expressed exactly in terms of $\eta$ functions and quasimodular forms/Eisenstein series. In all cases, unless stated otherwise, the evidence for the claimed equality is a match of $q$-series to very high orders using other established expressions for the index.

\subsection{\label{subsec:class_S_a1_experiment}Class \texorpdfstring{$\cal S$}{S} indices of type \texorpdfstring{$\mathfrak{a}_1$}{a1}}

The Schur index of the class $\mathcal S$ theory of type $\mathfrak{a}_1$ associated to a genus $g$ surface with $s$ punctures can be written in ``TQFT form'' \cite{Gadde:2011ik,Gadde:2011uv} as,
\begin{equation}\label{eq:a1_index_with_pochhammers}
    \mathcal I^{\mathfrak{a}_1}_{g,s}(q) = q^{\frac{c_{4d}}{2}}\;\frac{(q;q)_\infty^{2(g-1-s)}}{(1-q)^{2g-2+s}}\sum_{\lambda=0}^\infty \frac{(\lambda+1)^s}{{\rm dim}_q(\lambda)^{2g-2+s}}~.
\end{equation}
where the four-dimensional central charge is given in terms of the genus and number of punctures by
\begin{equation}
    c_{4d} = \frac{13(g-1)+5s}{6}~.
\end{equation}
This is also related to the partition function of $q$-deformed two-dimensional Yang-Mills theory in the zero-area limit. Based on the analysis of \cite{Beem:2017ooy} (see also the earlier \cite{Razamat:2012uv}), this index is expected to exhibit modular properties in this normalisation with the $q^{c_{4d}/2}$ prefactor, which matches the standard normalisation for the vacuum character of the associated VOA. The index in \eqref{eq:a1_index_with_pochhammers} can be re-written by converting the $q$-Pochhammer symbols into Dedekind $\eta$ functions, leading to the following,
\begin{equation}\label{eq:a1_index_with_dedekinds}
    \mathcal I^{\mathfrak{a}_1}_{g,s}(q) = \eta(\tau)^{2(g-1-s)}P_{g,s}^{(1)}(q)~,
\end{equation}
where $P_{g,s}^{(1)}(q)$ is now a series in either $q$ or $q^{1/2}$,
\begin{equation}
    P_{g,s}^{(1)}(q) = \left(\frac{q}{(1-q)^2}\right)^{g-1+\frac{s}{2}}\sum_{\lambda=0}^\infty\frac{(\lambda+1)^s}{{\rm dim}_q(\lambda)^{2g-2+s}}~.
\end{equation}

It is this ``$\eta$-stripped index'' that will be the main object of our interest in this paper; for small $q$ it can be seen to behave as $q^{g-1+\frac{s}{2}}$, so for physical class $\mathcal{S}$ theories this is always regular at $q=0$ and in general a number of terms in the $q$-expansion will vanish, with the first nonzero coefficient (that of $q^{g-1+\frac{s}{2}}$) being one.

In the case of the rank-one $\mathfrak{d}_4$ described in the introduction, we have
\begin{equation}
    P_{0,4}^{(1)}(q) = -12\;\mtbE_2(\tau)\mtbE_4(\tau)+42\;\mtbE_6(\tau)~.
\end{equation}
Looking beyond this example to other theories, we find that this result (with $P_{g,s}^{(1)}$ admitting an expression in terms of Eisenstein series) is massively generalised, and the $\eta$-stripped Schur index of type $\mathfrak{a}_1$ is apparently \emph{always} expressible as a finite-order polynomial in Eisenstein series $\mtbE_{2k}(\tau)$. For unpunctured ($s=0$ cases), the results look especially simple. To list a few examples, we find
\begin{equation}
\begin{aligned}
    P_{2,0}^{(1)}(q) &= \frac{1}{24}+\frac{\mtbE_2(\tau)}{2}~,\\
    P_{3,0}^{(1)}(q) &= -\frac{11}{1440}-\frac{\mtbE_2(\tau)}{12}+\frac{\mtbE_4(\tau)}{2}~,\\
    P_{4,0}^{(1)}(q) &= \frac{191}{120960}+\frac{\mtbE_2(\tau)}{60}-\frac{\mtbE_4(\tau)}{8}+\frac{\mtbE_6(\tau)}{2}~.\\
\end{aligned}
\end{equation}
These (and many more cases that can be easily checked at higher genus) are all quasimodular forms of mixed weight (with maximum weight $2g-2$) and depth one, with the additional (surprising) simplicity that they are just sums of Eisenstein series, with no more general products of Eisensteins. In other words, we have
\begin{equation}\label{eq:genus_zero_ansatz}
    P_{g,0}^{(1)}(q) = \sum_{i=0}^{g-1} \mathfrak a_i^{(g,0)} \mtbE_{2i}(\tau)~,
\end{equation}
for appropriate values of the coefficients $\mathfrak a_i^{(g,0)}$. Additionally, as the $q$-series for $P_{g,0}^{(1)}(q)$ must start with $q^{g-1}$ with unit coefficient, imposing the correct leading behaviour on \eqref{eq:genus_zero_ansatz} leads to $g$ constraints on the coefficients, exactly fixing them.

Amusingly, one may observe upon studying the solutions of these constraints that in general $\mathfrak a_i^{(g,0)}$ can be expressed as a polynomial in $g$ of order $(g-1-i)$. In fact, the coefficient $a_{g-1}^{(g,0)}$ is always equal to $1/2$ while the rest of the polynomial always have $g=1$ as a root. Interestingly, the polynomial for the coefficient of the constant term carries a minus sign relative to the rest of the coefficients for a fixed value of $g$. Therefore, we can write
\begin{equation}
\begin{aligned}
    \mathfrak a_{g-1}^{(g,0)} &= \frac12 ~,\quad \mathfrak a_{0}^{(g,0)} = -(g-1)\;\mathfrak f_{g-1}(g-1) ~, \\[0.2cm]
    \mathfrak a_i^{(g,0)} &= (g-1)\;\mathfrak f_{g-1-i}(g) ~,\quad {\rm for}\;0<i<g-1~,
\end{aligned}
\end{equation}
where $\mathfrak f_k(g)$ are the following polynomials of order $(k-1)$ in $g$; for example,
\begin{equation}
\begin{aligned}
    \mathfrak f_1(g) &= -\frac{1}{24}~,\quad \mathfrak f_2(g) = \frac{5g-4}{2880}~,\\[0.1cm]
    \mathfrak f_3(g) &= -\frac{35g^2-49g+18}{725760}~,\\[0.1cm]
    \mathfrak f_4(g) &= \frac{175g^3-315g^2+206g-48}{174182400}~,\\[0.1cm]
    \mathfrak f_5(g) &= -\frac{385g^4-770g^3+671g^2-286g+48}{22992076800}~.
\end{aligned}
\end{equation}
We have not identified a closed form for the general $\mathfrak{f}_k(g)$ polynomials.


The inclusion of punctures leads to more complicated polynomials of Eisenstein series, as well as twisted Eisenstein series, which arise for odd numbers of punctures. It turns out (nontrivially) that the polynomials appearing can in all examples be rewritten as derivatives of Eisenstein series (using Ramanujan identities). For example, we have,
\begin{equation}
\begin{aligned}
    P_{0,3}^{(1)}(q)    &= 24\,\mtbE_4\begin{bmatrix}-1\\+1\end{bmatrix}(\tau)+21\,\mtbE_4(\tau)~,\\[0.1cm]
    P_{0,4}^{(1)}(q)    &= 3\,\mtbE_4(\tau)^{(1)}~,\\[0.1cm]
    P_{1,1}^{(1)}(q)    &= \mtbE_2\begin{bmatrix}-1\\+1\end{bmatrix}(\tau)-\tfrac{1}{24}\mtbE_2(\tau)~,\\[0.1cm]
    P_{0,6}^{(1)}(q)    &= \tfrac{1}{2}\mtbE_4(\tau)^{(3)}-10\,\mtbE_6(\tau)^{(1)}~,\\[0.1cm]
    P_{2,1}^{(1)}(q)    &= \tfrac14\mtbE_2\begin{bmatrix}-1\\+1\end{bmatrix}(\tau)^{(1)}-\tfrac18\mtbE_2\begin{bmatrix}-1\\+1\end{bmatrix}(\tau)+\tfrac{1}{192}\mtbE_2(\tau)~.
\end{aligned}
\end{equation}
The question of which (derivatives of) Eisenstein series appear seems to have a general answer, and all results we have checked are consistent with the following conjectural Ansatz for the Eisenstein expression of a general class $\mathcal S$ index of type $\mathfrak{a}_1$.
{\small
\begin{equation}\label{a1conj}
\begin{aligned}
    P_{g,s}^{(1)}(q) = &\left[\sum_{i=0}^{g-1} {\mathfrak a}^{(g,s)}_{i}\,{\mtbE}_{2i}\begin{bmatrix}(-1)^s\\+1\end{bmatrix}(\tau)\right]^{(s)} \\
    \hspace*{-0.3cm}+\sum_{j=0}^{\left\lfloor\frac{s-1}2\right\rfloor}&\left[{\mathfrak b}^{(g,s)}_j \mtbE_{2j+2}(\tau)+{\mathfrak c}^{(g,s)}_j {\mtbE}_{2j+2}\begin{bmatrix}(-1)^s\\ +1\end{bmatrix}(\tau)\right]^{(s-2j-1)}~.
\end{aligned}
\end{equation}}\\[-10pt]
\noindent We impose ${\mathfrak b}_0^{(0,s)}={\mathfrak c}_0^{(0,s)}=0$, and the rest of the coefficients ${\mathfrak a}_i^{(g,s)}$, ${\mathfrak b}_j^{(g,s)}$, and ${\mathfrak c}_j^{(g,s)}$ are fixed by demanding that the first $g-1+\frac{s}{2}$ terms vanish, while the coefficient of $q^{g-1+\frac{s}{2}}$ is 1. This turns out to be a consistent set of requirements which can be uniquely solved. In the case $s=0$ this construction reduces to our previous discussion in terms of straight Eisenstein series (with no twists or derivatives).

\subsection{\label{subsec:a2_class_S}Class \texorpdfstring{$\cal S$}{S} indices of type \texorpdfstring{$\mathfrak a_2$}{a2}}

For class $\mathcal{S}$ theories of type $\mathfrak a_2$, the Schur index for a genus $g$ surface with $s$ maximal and $t$ minimal punctures is given in TQFT form as \cite{Gadde:2011ik,Gadde:2011uv}
\begin{widetext}
\begin{equation}\label{eq:a2_index_with_dedekinds}
\begin{aligned}
    \mathcal I_{g,s,t}^{\mathfrak a_2}(q) = q^{\frac{c_{4d}}{2}}\;\frac{(q;q)_\infty^{4(g-1)-6s}}{(q^{\frac12};q)_\infty^{2t}}
    \;&\frac{\left[(1-q)(1-q^{\frac12})\right]^{2t}}{\left[(1-q)^2(1-q^2)\right]^{2(g-1)+s+t}}
    \;\sum_{\lambda_1,\lambda_2=0}^\infty &\frac{{\rm dim}_{\mathfrak{su}(3)}(\lambda_1,\lambda_2)^{s}\chi^{\mathfrak{su}(3)}_{\lambda_1,\lambda_2}\left(q^{\frac12},q^{-\frac12},1\right)^{t}}{{\rm dim}_q(\lambda_1,\lambda_2)^{2g-2+s+t}}~,
\end{aligned}
\end{equation}
\end{widetext}
where now the four-dimensional central charge in given in terms of the topological data $g,s,t$ by
\begin{equation}
    c_{4d} = \frac{100(g-1)+42s+25t}{12}~.
\end{equation}
We will again be interested in the index stripped of $\eta$ functions. Since the index now contains $q$-Pochhammers with $q^{\frac12}$ argument, the $\eta$ prefactor contains powers of $\eta(\tau)$ as well as $\eta(\tau/2)$. We introduce the $\eta$-stripped index as follows,
\begin{equation}
\begin{aligned}
    \mathcal I_{g,s,t}^{\mathfrak a_2}(q) = \frac{\eta(\tau)^{4(g-1)+2t}}{\eta(\tau)^{6s}\eta(\tau/2)^{2t}} P^{(2)}_{g,s,t}(q)~.
\end{aligned}
\end{equation}
The stripped indices $P^{(2)}_{g,s,t}(q)$ again admit expressions as polynomials in $\mtbE_2(\tau)$ along with holomorphic modular forms over $\Gamma_1$ or $\Gamma^0(2)$ as appropriate, the latter corresponding to cases with half integer powers of $q$. Here we display a few representative examples,
\begin{equation}\label{eq:a2_quasimodular_indices_examples}
\begin{aligned}
    P^{(2)}_{2,0,0}(q) &= -\tfrac{47}{45360}-\tfrac{11\,\mtbE_2(\tau)}{720} - \tfrac{\mtbE_2(\tau)^2}{24} + \tfrac{\mtbE_4(\tau)}{24} + \tfrac{\mtbE_6(\tau)}{6}~,\\[0.2cm]
    P^{(2)}_{3,0,0}(q) &= \tfrac{1422991}{70053984000} + \tfrac{269\,\mtbE_2(\tau)}{950400} + \tfrac{\mtbE_2(\tau)^2}{1440} - \tfrac{5017\,\mtbE_4(\tau)}{3628800}~,\\[0.2cm]
    &\quad - \tfrac{\mtbE_2(\tau)\mtbE_4(\tau)}{280} + \tfrac{23\,\mtbE_6(\tau)}{12960} + \tfrac{\mtbE_4(\tau)^2}{120} - \tfrac{\mtbE_4(\tau)\mtbE_6(\tau)}{66}~,\\[0.2cm]
    &\quad - \tfrac{27\,\mtbE_4(\tau)^3}{1001} + \tfrac{35\,\mtbE_6(\tau)^2}{429}~,\\[0.2cm]
    P^{(2)}_{0,2,1}(q) &= \tfrac{\Theta_{2,2}(\tau)}{512}~,\\[0.2cm]
    P^{(2)}_{0,3,0}(q) &= \tfrac{864\,\mtbE_{10}^{(1)}(\tau)}{11} - \tfrac{\Delta(\tau)}{1050}~,\\[0.2cm]
    P^{(2)}_{0,1,3}(q) &= 3\,\mtbE_4^{(1)}(\tau)~,\\[0.2cm]
    P^{(2)}_{1,1,0}(q) &= -\tfrac{\mtbE_2(\tau)^3}{6}+\tfrac{\,\mtbE_2(\tau)\mtbE_4(\tau)}{2}+\tfrac{7\,\mtbE_6(\tau)}{6}~,\\[0.2cm]
    P^{(2)}_{1,0,1}(q) &= \tfrac{1}{24}+\tfrac{\mtbE_2(\tau)}{2}~,\\[0.2cm]
    P^{(2)}_{1,0,2}(q) &= \tfrac{1}{192}+\tfrac{\mtbE_2(\tau)}{24}-\tfrac{\Theta_{0,1}(\tau)}{144}+\tfrac{\mtbE_2(\tau)^2}{2}~,\\[0.2cm]
    &\hspace{1.5cm} -\tfrac{\mtbE_2(\tau)\Theta_{0,1}(\tau)}{24}+\tfrac{\Theta_{1,1}(\tau)}{288}-\tfrac{\Theta_{0,2}(\tau)}{576}~,
\end{aligned}
\end{equation}
where $\Delta(\tau)$ is the modular discriminant. For cases with $\Gamma^0(2)$ quasimodular expansions, we have used the $\Theta_{r,s}$ basis (as defined in the appendix) rather than one involving twisted Eisenstein series as a matter of computational convenience.

Some comments are in order about these results. Generically, the Eisenstein expressions comprise modular forms over $\Gamma_1$ only for $t=0$ (in this case, the associated VOA is expected to include only integer-weight states). The only two exceptions to this rule that we have found are displayed in \eqref{eq:a2_quasimodular_indices_examples}. The first of these is the sphere with one maximal and three minimal punctures, and the other is the genus one torus with one minimal puncture. These both admit expansions in modular forms over $\Gamma_1$ even though they have $t\neq0$. The full Schur indices in these cases still have half integer powers of $q$, but these accounted for entirely by the $\eta$ prefactor. Note that the theory with $(g,s,t)=(0,1,3)$ is equivalent to the $\mathfrak a_1$ theory with $(g,s)=(0,4)$ along with fields, while the $\mathfrak a_2$ theory with $(g,s,t)=(1,0,1)$ is $\mathfrak a_2$ ${\cal N}=4$ SYM with free fields, so the $q^{\frac12}$ dependence in these cases in fact comes entirely from free fields, which indeed contribute powers of $\eta(\tau/2)$ in the prefactor.

For these same two cases, the Eisenstein expressions are identical to type $\mathfrak a_1$ Eisenstein expressions associated to the four punctured sphere and unpunctured genus-two surface, respectively, \emph{i.e.},
\begin{equation}\label{eq:miraculous_equivalence_a1a2}
    P_{0,1,3}^{(2)}(q) = P_{0,4}^{(1)}(q) ~,\quad P_{1,0,1}^{(2)}(q) = P_{2,0}^{(1)}(q)~.
\end{equation}
The first of these is a consequence of the physical relationship we just mentioned, while the second is a more mysterious coincidence of $q$-series.

We do not have a simple conjecture for the $\eta$-stripped indices in this class of theories that is on par with \eqref{a1conj} for type $\mathfrak a_1$. However, in the next section we will still manage to propose a (less rigid) Ansatz that substantially simplifies the process of searching for Eisenstein expressions for these indices.

\subsection{\label{subsec:a3_class_s_results}Class \texorpdfstring{$\cal S$}{S} indices of type \texorpdfstring{$\mathfrak a_3$}{a3}}

Here we wish to briefly touch on the class $\mathcal S$ indices of type $\mathfrak a_3$, restricting to the case of all maximal punctures. The $q$-Pochhammer symbols in this case can all be converted into factors of $\eta(\tau)$ and we define the $\eta$-stripped indices by
\begin{equation}
    \mathcal I^{\mathfrak a_3}_{g,s} (q) = \eta(\tau)^{6(g-1)-12s} P^{(3)}_{g,s}(q)~.
\end{equation}
We again find that the $\eta$-stripped indices admit mixed-weight quasimodular expressions, though in general they are quite lengthy. To keep things succinct, we present only address the cases of the unpunctured genus-two theory and the trinion theory (three-punctured sphere),
\begin{equation}
\begin{aligned}
    \mathcal I^{\mathfrak a_3}_{2,0} (q) &= \eta(\tau)^6 P^{(3)}_{2,0}(q) ~,\\
    \mathcal I^{\mathfrak a_3}_{0,3} (q) &= \eta(\tau)^{-42} P^{(3)}_{0,3}(q)~.
\end{aligned}
\end{equation}
We find the following mixed-weight quasimodular expressions,
\begin{widetext}
\begin{equation}\label{a3indices}
\begin{aligned}
    P^{(3)}_{2,0} (q) &= -\frac{1326517}{72648576000} - \frac{70093\,\mtbE_2(\tau)}{239500800} - \frac{97\,\mtbE_2(\tau)^2}{86400}
    + \frac{13799\,\mtbE_4(\tau)}{14515200} - \frac{\mtbE_2(\tau)^3}{864} + \frac{13\,\mtbE_2(\tau)\mtbE_4(\tau)}{5040}\\[0.2cm]
    &\hspace{4cm}+ \frac{127\,\mtbE_6(\tau)}{120960} + \frac{\mtbE_2(\tau)\mtbE_6(\tau)}{180} - \frac{3\,\mtbE_4(\tau)^2}{1120}
    - \frac{\mtbE_4(\tau)\mtbE_6(\tau)}{264} + \frac{29\,\mtbE_4(\tau)^3}{4004} - \frac{9\,\mtbE_6(\tau)^2}{572}~,\\[0.2cm]
    P^{(3)}_{0,3}(q) &= \frac{\mtbE_2(\tau)^3}{442368}\Big[2\,\Theta_{2,7}(\tau)-15\,\Theta_{3,6}(\tau)-307\,\Theta_{4,5}(\tau)\Big]
    +\frac{\mtbE_2(\tau)^2}{884736}\Big[\Theta_{2,8}(\tau)-8\,\Theta_{3,7}(\tau)+52\,\Theta_{4,6}(\tau)+115
    \,\Theta_{5,5}(\tau)\Big]\\[0.2cm]
    &+\frac{\mtbE_2(\tau)}{106274488320}\Big[12096\,\Theta_{2,8}(\tau)
    +10010\,\Theta_{2,9}(\tau)
    -103680\,\Theta_{3,7}(\tau)-85085\,\Theta_{3,8}(\tau)+1023930\,\Theta_{4,6}(\tau)\\[0.2cm]
    &\hspace{4cm}+350350\,\Theta_{4,7}(\tau)+1014966\,\Theta_{5,5}(\tau)-1876875\,\Theta_{5,6}(\tau)\Big]~,\\[0.2cm]
    &+\frac{1}{3825881579520}\Big[36288\,\Theta_{2,9}(\tau)+10010\,\Theta_{2,10}(\tau)
    -329184\,\Theta_{3,8}(\tau)-90090\,\Theta_{3,9}(\tau)+788076\,\Theta_{4,7}(\tau)\\[0.2cm]
    &\hspace{4cm}+385385\,\Theta_{4,8}(\tau)-6337116\,\Theta_{5,6}(\tau)-1001000\,\Theta_{5,7}(\tau)+2297295\,\Theta_{6,6}(\tau)\Big]~.
\end{aligned}
\end{equation}
\end{widetext}
We remark that whereas for the three-punctured sphere in types $\mathfrak{a}_1$ and $\mathfrak{a}_2$ the index was of definite modular weight, here we have the $\mathfrak{a}_3$ trinion index already being a mixed-weight quasimodular form.

\subsection{\label{subsec:n4_results}\texorpdfstring{$SU(N)$}{SU(N)} \texorpdfstring{$\mathcal N=4$}{N=4} super Yang--Mills}

The Schur indices of $\mathcal N=4$ super Yang--Mills theory with gauge algebra $\mathfrak{su}(n)$ were obtained in closed form in \cite{Bourdier:2015wda}. The results in that work were given in terms of the \emph{complete elliptic integrals} of the first and second kind. These complete elliptic integrals can themselves be understood as hypergeometric functions with argument given by the \emph{elliptic modulus} $\kappa(\tau)$,
\begin{equation}
\begin{aligned}
    K(\kappa) &= \frac\pi2\;{}_2F_1\left(\frac12\,,\,+\frac12\,,\,1\,,\,\kappa(\tau)^2\right)~,\\
    E(\kappa) &= \frac\pi2\;{}_2F_1\left(\frac12\,,\,-\frac12\,,\,1\,,\,\kappa(\tau)^2\right)~,
\end{aligned}
\end{equation}
where $\kappa(\tau) = \theta_2(\tau)^2/\theta_3(\tau)^2$. From a modular perspective, these can be expressed in a simpler form in terms of the second Eisenstein series and theta functions. In particular, the elliptic integral of the first kind is just a power of a Jacobi theta function,
\begin{equation}
    K(\kappa) = \frac\pi2 \theta_3(\tau)^2~,
\end{equation}
while using the relation $\pi^{-2}K(K-E)=\Theta_{0,1}/12 + \mtbE_2$, we also have the equality for the complete elliptic integral of the second kind,
\begin{equation}
    E(\kappa) = \frac\pi2\left(\frac{2\;\theta_3(\tau)^2-\theta_2(\tau)^2-12\;\mtbE_2(\tau)}{3\;\theta_3(\tau)^2}\right)~.
\end{equation}
With these relations in hand, the results of \cite{Bourdier:2015wda} can be cast in the format we are promoting in the present work. It works out that the Dedekind $\eta$ factor just depends on whether the rank of the gauge algebra is even or odd. In particular, we have the following expression for the index of $\mathcal N=4$ SYM with gauge algebra $\mathfrak a_{n}$,
\begin{equation}
    \mathcal I_{\mathfrak a_n}(q) = \left[\frac{\eta(\tau/2)^2}{\eta(\tau)^4}\right]^{n({\rm mod}\;2)} P^{\mathcal{N}=4}_{\mathfrak a_n}(q)~.
\end{equation}
As in the class $\mathcal S$ examples above, for even $n$ (where the index is a $q$ series) we find an expansion in $\Gamma_1$ quasimodular forms. Below we display the first few such indices explicitly---it is a simple matter to extend to higher rank.
\begin{widetext}
\begin{equation}
\begin{aligned}
    P^{\mathcal{N}=4}_{\mathfrak a_2}(q) &= \frac{1}{24}+\frac{\mtbE_2(\tau)}{2}~,\qquad\qquad\qquad
    P^{\mathcal{N}=4}_{\mathfrak a_4}(q) = \frac{3}{640}+\frac{\mtbE_2(\tau)}{16}+\frac{1}{4}\left[\frac{\mtbE_2(\tau)^2}{2}-\;\mtbE_4(\tau)\right]~,\\[0.3cm]
    P^{\mathcal{N}=4}_{\mathfrak a_6}(q) &= \frac{5}{7618}+\frac{37\;\mtbE_2(\tau)}{3840}+\frac{5}{96}\left[\frac{\mtbE_2(\tau)^2}{2}-\mtbE_4(\tau)\right]+\frac{1}{48}\left[\mtbE_2(\tau)^3-6\;\mtbE_2(\tau)\mtbE_4(\tau)+8\;\mtbE_6(\tau)\right]~.
\end{aligned}
\end{equation}
\end{widetext}
For odd $n$, the series includes half-integer powers of $q$ (even after removing $\eta$ factors). The expansion is then in quasimodular forms over the $\Gamma^0(2)$, and again we display the first few indices explicitly,
\begin{widetext}
\begin{equation}
\begin{aligned}
    P^{\mathcal{N}=4}_{\mathfrak a_1}(q) &= \frac{\mtbE_2(\tau)}{2}+\frac{\Theta_{0,1}(\tau)}{24}~,\\[0.3cm]
    P^{\mathcal{N}=4}_{\mathfrak a_3}(q) &= \frac{\mtbE_2(\tau)^2}{8}+\frac{\mtbE_2(\tau)\Theta_{0,1}(\tau)}{48}+\frac{1}{576}\left[\frac{\Theta_{0,2}(\tau)}{2}-\Theta_{1,1}(\tau)\right]+\frac{1}{12}\left[\frac{\mtbE_2(\tau)}{2}+\frac{\Theta_{0,1}(\tau)}{24}\right]~, \\[0.3cm]
    P^{\mathcal{N}=4}_{\mathfrak a_5}(q) &= \frac{\mtbE_2(\tau)^3}{16}+ \frac{\mtbE_2(\tau)}{2}\left[ \frac{\mtbE_2(\tau)^2}{8}+ \frac{\mtbE_2(\tau)\Theta_{0,1}(\tau)}{48}+ \frac{1}{576}\left[ \frac{\Theta_{0,2}(\tau)}{2}-\Theta_{1,1}(\tau)\right]\right]+ \frac{1}{27648}\left[ \frac{\Theta_{0,3}(\tau)}{3}- \frac{\Theta_{1,2}(\tau)}{5}\right]\\[0.3cm] 
    &\hspace{1cm}~+\frac{\mtbE_2(\tau)}{24}\left[ \frac{\mtbE_2(\tau)}{2}+ \frac{\Theta_{0,1}(\tau)}{24}\right]+ \frac{\mtbE_2(\tau)\Theta_{0,1}(\tau)}{24}+ \frac{1}{3456}\left[ \frac{\Theta_{0,2}(\tau)}{2}-\Theta_{1,1}(\tau)\right]+ \frac{1}{90}\left[ \frac{\mtbE_2(\tau)}{2}+ \frac{\Theta_{0,1}(\tau)}{24}\right]~.
\end{aligned}
\end{equation}
\end{widetext}
Of course, these are actually equivalent to class $\mathcal{S}$ Schur indices of type $\mathfrak a_n$ for the torus with one minimal puncture (up to an extra free hypermultiplet). A free hypermultiplet only contributes $\eta$ prefactors to the Schur index, so they will have the same quasimodular expansions.

\section{\label{sec:structure_and_lessons}General structure of the results and general lessons}

The results of the previous section exhibit a great deal of commonality, with the Schur indices all admitting expressions in terms of Eisenstein series (including the $\mtbE_2(\tau)$ series) and, in the case with half-integer powers of $q$, $\Gamma^0(2)$ modular forms $\Theta_{r,s}(\tau)$. These are all examples of \emph{quasimodular forms}. In this section, we review some basic structure theory of these functions and comment on the picture that emerges for Schur indices in terms of quasimodular forms.

\subsection{\label{subsec:quasimodular_mlde}Quasimodular forms and MLDE}

A quasimodular form of weight $k$ and depth $\leqslant p$ on $\Gamma$ is defined to be a function $\phi(\tau)$ on which the weight $k$ modular action $(c\tau+d)^{-k}\phi(\gamma\tau)$ generates a polynomial of degree $\leqslant p$ in $\frac{c}{c\tau+d}$ with coefficients that are holomorphic functions on the upper half plane. For a pedagogical review of the subject, see \cite{ZagierBook:2008}. It can then be proven that every quasimodular form of weight $k$ and depth $\leqslant p$ on $\Gamma$ (being either the full modular group or a congruence subgroup), can be written as \cite{KanekoZagier},
\begin{equation}
    \phi(\tau) = \sum_{r=0}^p \mathbb E_2(q)^r \phi_r(q)~,\qquad\phi_r\in M_{k-2r}(\Gamma)~,
\end{equation}
where the property of transforming into a polynomial of degree at most $p$ in $\frac{c}{c\tau+d}$ imposes the restriction to at most $p$ powers of the anomalous Eisenstein series $\mtbE_2(\tau)$.

More simply stated, the ring of all quasimodular forms over $\Gamma_1$ is a graded filtered ring, finitely and freely generated over $\mathbb C$ by generators $\mtbE_2(\tau)$, $\mtbE_4(\tau)$, and $\mtbE_6(\tau)$. We denote by $\widetilde{M}_k$ the vector space of $\Gamma_1$-quasimodular forms of weight $k$ and by $\widetilde M_k^{(\leqslant p)}$ the subspace with at most $p$ powers of $\mtbE_2(\tau)$. With respect to these we have,
\begin{equation}
    \mathbb C[\mtbE_2,\mtbE_4,\mtbE_6] = \widetilde M_* = \bigoplus_k \widetilde M_k = \bigoplus_k \bigcup_p \widetilde M_k^{(\leqslant p)}~.
\end{equation}
For $\Gamma^0(2)$, quasimodular forms are again polynomials in $\mtbE_2(\tau)$ with coefficients being the functions $\Theta_{r,s}(q)$ defined in the appendix. An important feature of the ring of quasimodular forms is that it is closed under differentiation, and this respects the filtration by depth,
\begin{equation}
    D\left(\widetilde M_k^{(\leqslant p)}\right) \subset \widetilde M_{k+2}^{(\leqslant p+1)} ~,\quad D\colonequals q\partial_q~.
\end{equation}
In the case of $\Gamma_1$-quasimodular forms, this follows very simply from the famed Ramanujan identities for the derivatives of Eisenstein series, but the same can be shown to hold true for more general $\Gamma$.

We are interested in quasimodular forms that arise in the context of studying solutions to finite-order modular linear differential equations (acting at weight zero). Generally speaking, the Frobenius solutions of a MLDE are $q$-series, potentially with finitely many powers of $\log q$ appearing, and these transform into linear combinations of each other under modular transformations. From the definition of quasimodular forms, for $\phi(\tau)$ quasimodular of weight $k$ and depth $p$, the modular transformed $(c\tau+d)^{-k}\phi(\gamma\tau)$ will have terms that are $q$ series with coefficients of the form $(c\tau+d)^{-m}$ for $m\leqslant p$. Therefore, if we instead consider the \emph{weight $k-p$} action on $\phi$, we obtain a polynomial of degree at most $p$ in $c(c\log(q)+d)$ with quasimodular coefficients. This motivates treating quasimodular forms, as in the discussion in the introduction of this paper, as weight $k-p$ objects. In particular, this means that if we take the combination $\eta(\tau)^{2(p-k)}\phi(\tau)$ then this will naturally transform as a weight-zero logarithmic vector-valued modular function. An explicit analysis of this transformation shows that this will in fact be a component of a $p+1$-dimensional modular vector of functions. Further multiplying $\eta(q)^{2(p-k)}\phi(q)$ by additional powers of $\eta(q)^2$ then further increases the size of the modular vector by one a piece.

In the case of conventional modularity, it is well-known that a $d$-dimensional vector-valued modular form is the solution of an order $d$ modular linear differential equation. In the logarithmic/quasimodular case, we are not aware of an analogously strong result, but we find that $d$-dimensional logarithmic vector-valued modular functions arising in the context of Schur indices are indeed solutions of MLDEs of order $d$, so we will assume in what follows that this should be the case more generally. It would be interesting to establish such a result systematically.

\subsection{\label{subsec:schur_quasimodular}Schur indices and quasimodular forms}

In the examples of the previous section, we encountered mixed-weight quasimodular objects with a common Dedekind $\eta$ factor. Each (fixed-weight) quasimodular form in such an expression can, as above, be thought of as giving an element of a logarithmic, vector-valued modular function, so the mixed-weight objects will be vectors in the modular representation obtained as the direct sum of the fixed-weight representations. This then proves useful for understanding the sorts of MLDEs that should be expected for these indices.

Let us consider the simplest example of this, the genus-two $\mathfrak{a}_1$ theory with no punctures\footnote{See \cite{Kiyoshige:2020uqz,Beem:2021jnm} for recent investigations of this theory from a VOA perspective.}, with Schur index $\mathcal I_{2,0}^{\mathfrak a_1}(q) = \eta(\tau)^2P_{2,0}^{(1)}(q)$, where
\begin{equation}\label{eq:genus_two_a1_mixed_weight}
    P_{2,0}^{(1)}(q) = \frac{1}{24}+\frac{\mtbE_2(\tau)}{2}~.
\end{equation}
We have the sum of two quasimodular forms of weights zero and two, with depths zero and one, respectively. Per the discussion in the previous subsection, we can think of the constant function as lying in a one dimensional modular vector and $\eta(\tau)^{-2}\mtbE_2(\tau)$ as an element of a two-dimensional (logarithmic) modular vector, both of weight zero. Indeed,
\begin{equation}
    \frac{\mtbE_2(\gamma\tau)}{\eta(\gamma\tau)^2} \;\sim\; \frac{c}{2\pi i}\left(\frac{(\log q)\mtbE_2(\tau)+1}{\eta(\tau)^2}\right) + d\;\frac{\mtbE_2(\tau)}{\eta(\tau)^2}~,
\end{equation}
where we have ignored an overall phase contributed by the modular transformation of the Dedekind $\eta$ function. The index involves extra factors of $\eta(\tau)^2$, whose effect will be to introduce additional powers of $(c\tau+d)$ to the modular transformation of the index. As explained above, multiplying by $\eta(\tau)^2$ will increase the size of the modular vector in which a given piece of the index transforms, and so will increase the order of the MLDE that it satisfies. We then have that $\eta(\tau)^2(1)$ lies in a two-dimensional modular vector (and satisfies an order-two MLDE) while $\eta(\tau)^{2}\mtbE_2(\tau)$ by $\eta(\tau)^4$ lies in a four-dimensional modular vector and satisfies an order-four MLDE. The index lies in the direct sum of these two modular representations, so the Schur as a whole lives in a six-dimensional logarithmic vector-valued modular function. This is, indeed, in perfect agreement with the fact that $\mathcal I_{2,0}^{\mathfrak a_1}(q)$ satisfies a sixth-order MLDE \cite{Beem:2017ooy}.

To represent this situation at the level of modular vectors, we write,
\begin{equation}
    \mathcal I_{2,0}^{\mathfrak a_1}(q) \in \eta(\tau)^2 \left[\widetilde M_0^{(\leqslant0)}(\Gamma_1)\oplus\widetilde M_2^{(\leqslant1)}(\Gamma_1)\right]~.
\end{equation}
The exact numerical coefficients, equivalently the specific element of the above space of quasimodular forms, can be determined by demanding the correct behaviour as $q\rightarrow0$. In this case, the $q$-series for the elements of each fixed-weight space of quasimodular forms starts with $1$, and the correct combination to yield the true Schur index is to cancel the leading constant term and fix the coefficient of the $O(q)$ term to one. This means that once the summands of the (decomposable) vector-valued modular function is fixed, the Schur index is identified with the vector that is the most regular at $q=0$. This is a pattern that persists in all other examples we have considered.

In general, there are some constraints on the sorts of quasimodular forms that can appear in the Eisenstein expansion of a given Schur index, and this is practically useful when attempting to identify Eisenstein expressions by matching $q$-series. For simplicity, let us restrict to cases with pure $\eta(\tau)$ prefactors (so without twisted punctures in type $\mathfrak{a}_2$. We then have indices of the form
\begin{equation}\label{General_index_eta}
    \mathcal I(q) = \eta(\tau)^{2\rho} P\left(\{\mtbE\}\right)~,
\end{equation}
where $\rho$ is a general power depending on the Schur index\footnote{In the present setting, we have an opinion regarding the appropriate power of $\eta$ functions to strip off based on the TQFT expression for the class $\mathcal{S}$ Schur index. More generally, it is an interesting question whether the appropriate $\eta$ prefactor can be predicted \emph{a priori}.} and $P\left(\{\mtbE\}\right)$ is generally a sum of quasimodular forms. When $\rho\geqslant0$, the Eisenstein series appearing in $P\left(\{\mtbE\}\right)$ are in principle unconstrained, but for $\rho\leq0$ this will not necessarily transform as a logarithmic vector-valued modular function.

Indeed, we previously observed that a quasimodular form of weight $k$ and depth $p$ can be thought of as a logarithmic vector-valued modular form of weight $k-p$ (or more generally, of weight $w\leqslant k-p$, as reducing the weight of the modular action only introduces extra logarithmic structure). Thus, for a fixed value of $\rho$, the quasimodular forms appearing must obey the constraint $k-p+\rho\geqslant0$, which becomes a meaningful constraint for large negative $\rho$ (see, \emph{e.g.}, $P_{0,3}^{(3)}(q)$ in \eqref{a3indices}). Indeed, this provides important simplifications to the Ans\"atze that one should use when looking for Eisenstein expressions via matching $q$-series.

In the following, we consider the classes of theories from the previous section from the perspective of the spaces of quasimodular forms they inhabit. We will find that even though a simple Ansatz for the Eisenstein expression for these indices is not always forthcoming, we do find (conjectural) Ans\"atze for the relevant quasimodular form spaces. Once we have determined that, we can follow the same procedure used above for the genus-two $\mathfrak{a}_1$ theory to predict the dimension of the modular vector in which the Schur index lives, and so the order of the MLDE that the Schur index satisfies.\footnote{Note that this computation typically gives the order of the MLDE of minimal order, which can generally have nonzero Wronskian index rather than being monic.}

\subsection{\label{subsec:class_S_a1_quasimodualr}Class \texorpdfstring{$\cal S$}{S} of type \texorpdfstring{$\mathfrak{a}_1$}{a1}}

The Eisenstein expressions for the $\mathfrak{a}_1$ theories contain derivatives of Eisenstein and twisted Eisenstein series. These are straightforward to identify as quasimodular forms over the full modular group and $\Gamma^0(2)$, respectively. To be precise, we have
\begin{equation}
    \mtbE_n(q)^{(m)} \;\in\; \widetilde M_{n+2m}^{(\leqslant m+\delta_{n,2})}(\Gamma_1)~,
\end{equation}
as well as an analogous expression for the twisted Eisenstein series and $\Gamma^0(2)$. For $n=0$, the $m^{\rm th}$ derivative always ends up in $\widetilde M_0^0$ whereas for $n=2$, as the depth is already one, after $m$ derivatives it is equal to $m+1$.

The conjectured form \eqref{a1conj} for these indices then implies that the $\eta$-stripped Schur indices for $\mathfrak a_1$ class $\mathcal{S}$ theories lie in the following direct sum of spaces of quasimodular forms,
\begin{equation}\label{a1_qmfspace}
\begin{aligned}
    P^{(1)}_{g,s} \;\in\; \tilde\delta_{s,0}\widetilde M_0^{(\leqslant0)}(\Gamma)\oplus\bigoplus_{i=1}^{g-1}\widetilde M_{2s+2i}^{(\leqslant s+\delta_{i,1})}(\Gamma)& \\
    \oplus\bigoplus_{j=0}^{\left\lfloor\frac{s-1}{2}\right\rfloor}\widetilde M_{2s-2j}^{(\leqslant s-1-2j+\delta_{j,1})}(\Gamma)&~,
\end{aligned}
\end{equation}
where $\Gamma$ is equal to $\Gamma_1$ or $\Gamma^0(2)$ for the number of punctures $s$ being even or odd, respectively, and we have introduced the notation $\tilde\delta_{s,0}=1-\delta_{s,0}$. From this we can determine the order of the MLDE that the Schur indices of type $\mathfrak a_1$ should satisfy. We first note that the power of the Dedekind eta factor is $\rho = g-s-1$. Focusing on terms in the first direct sum of \eqref{a1_qmfspace}, we see that for any fixed $i$, these will transform in $2s+2i+1$-dimensional modular vectors under weight-zero modular transformations. Including the prefactor, we get $s+g+2i$-dimensional vectors.

Analogously, the terms in the second direct sum correspond to modular vectors of dimension $g+s-2j$. Summing over $i$ and $j$ gives us the following predicted order for the MLDE that these Schur indices must satisfy,
\begin{equation}\label{a1_order}
\begin{aligned}
    {\rm Ord}^{\mathfrak a_1}_{g,s} &= (1-\delta_{g,0})\;g(2g+s-1) \\
    +\;(1-\delta_{s,0})&\;\left\lfloor\frac{s-1}{2}\right\rfloor\left(g+s-1-\left\lfloor\frac{s-1}{2}\right\rfloor\right)~.
\end{aligned}
\end{equation}
Comparing to \cite{Beem:2017ooy}, this reproduces the results found there up to the added feature that where the MLDEs of \cite{Beem:2017ooy} had irrational roots, our formula gives the number of rational roots (with multiplicity) only. This is because for these theories, the minimal MLDE has positive Wronskian index.

For example, for the case of the eight punctured sphere, \eqref{a1_order} predicts an MLDE of order twelve. This can be explicitly verified, and it turns out that the MLDE in questions has Wronskian index, $\ell=18$. All the indicial roots of this MLDE are rational and coincide with the rational roots observed in \cite{Beem:2017ooy}, but the MLDE in that work was monic of order sixteen with four irrational roots.

\subsection{\label{subsec:higher_rank_quasimodular}Higher rank class \texorpdfstring{$\cal S$}{S} and \texorpdfstring{$\mathcal{N}=4$}{N=4}}

From the explicit expressions for the Schur indices for class $\cal S$ theories of type $\mathfrak a_2$, we can read off the following spaces of quasimodular forms in which the relevant Eisenstein expressions reside,
\begin{equation}
\begin{aligned}
    P_{2,0,0}^{(2)} \;&\in\; \widetilde M_0^{(\leqslant0)}\oplus\widetilde M_2^{(\leqslant1)}\oplus\widetilde M_4^{(\leqslant2)}\oplus\widetilde M_6^{(\leqslant0)}~,\\
    P_{3,0,0}^{(2)} \;&\in\; \widetilde M_0^{(\leqslant0)}\oplus\widetilde M_2^{(\leqslant1)}\oplus\widetilde M_4^{(\leqslant2)}\oplus\widetilde M_6^{(\leqslant1)}\\
    &\hspace*{1.5cm}\oplus\widetilde M_8^{(\leqslant0)}\oplus\widetilde M_{10}^{(\leqslant0)}\oplus\widetilde M_{12}^{(\leqslant0)} ~,\\
    P_{0,2,1}^{(2)} \;&\in\; \widetilde M_8^{(\leqslant0)}~,\\
    P_{0,3,0}^{(2)} \;&\in\; \widetilde M_{12}^{(\leqslant1)}~,\\
    P_{0,1,3}^{(2)} \;&\in\; \widetilde M_6^{(\leqslant1)}~,\\
    P_{1,1,0}^{(2)} \;&\in\; \widetilde M_6^{(\leqslant3)}~,\\
    P_{1,0,1}^{(2)} \;&\in\; \widetilde M_0^{(\leqslant0)}\oplus\widetilde M_2^{(\leqslant1)}~,\\
    P_{1,0,2}^{(2)} \;&\in\; \widetilde M_0^{(\leqslant0)}\oplus\widetilde M_2^{(\leqslant1)}\oplus\widetilde M_4^{(\leqslant2)}\oplus\widetilde M_6^{(\leqslant0)}~.
\end{aligned}
\end{equation}
For $t=0$ these should be understood as quasimodular forms for $\Gamma_1$, and otherwise they are quasimodular forms over $\Gamma^0(2)$ apart from the exceptions that were noted earlier.

Though we do not have a construction analogous to \eqref{a1conj} for the Schur indices of higher rank, we can nevertheless extract from examples a conjecture for the space of quasimodular forms in which the Eisenstein expressions of type $\mathfrak a_2$ should sit in general. We have arrived at this conjecture by explicit calculations, with some guidance coming from the constraints associated with Dedekind $\eta$ prefactors when $\rho<0$ in \eqref{General_index_eta}. Before writing down the full expression, we consider some simplified subcases.

Let us first consider the $\mathfrak a_2$ theories with genus $g$ and no punctures. Recall that the Schur index for this case can be expressed as follows,
\begin{equation}
    \mathcal I_{g,0,0}^{\mathfrak a_2}(q) = \eta(q)^{4(g-1)}P_{g,0,0}^{(2)}(q)~,
\end{equation}
The quasimodular forms appearing in the indices for these theories can be seen (empirically) to fit into three categories: (1) maximum depth for a given weight, (2) depth one, and (3) depth zero, with the different cases occurring for different values of the weight. All in all the structure neatly generalises to the following expression for general values of $g$,
\begin{equation}\label{a2_genus_nopunc}
    P_{g,0,0}^{(2)} \;\in\; \bigoplus_{i=0}^2 \widetilde M_{2i}^{(\leqslant i)}\oplus\bigoplus_{j=3}^g \widetilde M_{2j}^{(\leqslant1)}\oplus\bigoplus_{k=g+1}^{3g-3} \widetilde M_{2k}^{(\leqslant0)}~.
\end{equation}
We will see that a version of this observation is true even after adding maximal and minimal punctures to the mix.

Next we look at the $\mathfrak a_2$ theories associated to spheres with all maximal punctures. A few representative examples are as follows,
\begin{equation}
\begin{aligned}
    P_{0,3,0}^{(2)} \;&\in\; \widetilde M_{12}^{(\leqslant1)}~,\\
    P_{0,4,0}^{(2)} \;&\in\; \widetilde M_{16}^{(\leqslant2)}\oplus\widetilde M_{18}^{(\leqslant4)}~,\\
    P_{0,5,0}^{(2)} \;&\in\; \widetilde M_{18}^{(\leqslant1)}\oplus\widetilde M_{20}^{(\leqslant3)}\oplus\widetilde M_{22}^{(\leqslant5)}\oplus\widetilde M_{24}^{(\leqslant7)}~,\\
    P_{0,6,0}^{(2)} \;&\in\; \widetilde M_{22}^{(\leqslant2)}\oplus\widetilde M_{24}^{(\leqslant4)}\oplus\widetilde M_{26}^{(\leqslant6)}\oplus\widetilde M_{28}^{(\leqslant8)}\oplus\widetilde M_{30}^{(\leqslant10)}~,
\end{aligned}
\end{equation}
where all quasimodular forms are over the full modular group $\Gamma_1$. In this case, the depths of the weight $k$ quasimodular forms are simply given by $k-3s-2$, and this can be generalised to the following conjectural decomposition,
\begin{equation}
    P_{0,s,0}^{(2)} \;\in\; \bigoplus_{i=0}^{\frac{3s+s({\rm mod}\,2)}{2}-5} \widetilde M_{3s+4+2i-s({\rm mod}\,2)}^{(\leqslant2i+2-s({\rm mod}\,2))}~.
\end{equation}

We now introduce minimal punctures to the above expression. We have the following examples,
\begin{equation}
{\small
\begin{aligned}
    P_{0,3,0}^{(2)} \,&\in\, \widetilde M_{12}^{(\leqslant1)}~,\\
    P_{0,3,1}^{(2)} \,&\in\, \widetilde M_{12}^{(\leqslant1)}\oplus\widetilde M_{14}^{(\leqslant3)}~,\\
    P_{0,3,2}^{(2)} \,&\in\, \widetilde M_{12}^{(\leqslant1)}\oplus\widetilde M_{14}^{(\leqslant3)}\oplus\widetilde M_{16}^{(\leqslant5)}~,\\
    P_{0,4,0}^{(2)} \,&\in\, \widetilde M_{16}^{(\leqslant2)}\oplus\widetilde M_{18}^{(\leqslant4)}~,\\
    P_{0,4,1}^{(2)} \,&\in\, \widetilde M_{14}^{(\leqslant0)}\oplus\widetilde M_{16}^{(\leqslant2)}\oplus\widetilde M_{18}^{(\leqslant4)}\oplus\widetilde M_{20}^{(\leqslant6)}~,\\
    P_{0,4,2}^{(2)} \,&\in\, \widetilde M_{14}^{(\leqslant0)}\oplus\widetilde M_{16}^{(\leqslant2)}\oplus\widetilde M_{18}^{(\leqslant4)}\oplus\widetilde M_{20}^{(\leqslant6)}\oplus\widetilde M_{22}^{(\leqslant8)}~.\\
\end{aligned}}
\end{equation}
We see that for an odd number of maximal punctures, increasing the number of minimal punctures by one amounts to including an additional quasimodular form of weight two greater than the previous largest one. However, the case with $t=0$ behaves slightly differently, in that a quasimodular form of weight $3s+2$ is absent from the sum. 

Upon increasing the number of minimal punctures for a fixed number of maximal punctures, the depths corresponding to a set of large weights turns out to be first $6s+1$ and then $6s$. This is the same pattern noted previously for $s=t=0$ where for large enough weight for given $g$, the quasimodular forms had depths $1$ and then $0$. We therefore propose that the general Eisenstein expression associated to punctured spheres lies in the following direct sum of spaces of quasimodular forms,
\begin{equation}\label{a2_max_min}
{\small
\begin{aligned}
    P_{0,s,t}& \;\in\; \bigoplus_{i=0}^{min\left(t-4+\frac{3s+\Lambda_0}{2},\;s+\left\lfloor\frac{s}{2}\right\rfloor+1\right)} \widetilde M_{3s+2-\Lambda_0+2i}^{(\leqslant2i-\Lambda_0)} \\
    & \bigoplus_{j=3}^{\left\lfloor\frac{t}{2}\right\rfloor} \widetilde M_{6s+2j}^{(\leqslant3s+1)}
    \bigoplus_{k=1}^{t-3-\left\lfloor\frac{t}{2}\right\rfloor} \widetilde M_{6s+2\left\lfloor\frac{t}{2}\right\rfloor+2k}^{(\leqslant3s)}~,
\end{aligned}}
\end{equation}
where we have defined
\begin{equation}
\Lambda_0 = -
\begin{cases}
    2-s({\rm mod}\;2) &,\; t=0 \\[0.2cm]
    \quad s({\rm mod}\;2) &,\; t\neq0~.
\end{cases}
\end{equation}

Finally, we move on to the case with arbitrary genus and no maximal punctures (but any number of minimal punctures). As before, adding minimal punctures for a fixed genus and number of maximal punctures adds a quasimodular form with the next allowed weight. For instance, we have the following examples
\begin{equation}
{\small
\begin{aligned}
    P_{1,0,1}^{(2)} &\in \widetilde M_0^{(\leqslant0)}\oplus\widetilde M_2^{(\leqslant1)}~,\\
    P_{1,0,2}^{(2)} &\in \widetilde M_0^{(\leqslant0)}\oplus\widetilde M_2^{(\leqslant1)}\oplus\widetilde M_4^{(\leqslant2)}~,\\
    P_{1,0,3}^{(2)} &\in \widetilde M_0^{(\leqslant0)}\oplus\widetilde M_2^{(\leqslant1)}\oplus\widetilde M_4^{(\leqslant2)}\oplus\widetilde M_6^{(\leqslant0)}~, \\
    P_{2,0,1}^{(2)} &\in \widetilde M_0^{(\leqslant0)}\oplus\widetilde M_2^{(\leqslant1)}\oplus\widetilde M_4^{(\leqslant2)}\oplus\widetilde M_6^{(\leqslant0)}\oplus\widetilde M_8^{(\leqslant0)}~,\\
    P_{2,0,1}^{(2)} &\in \widetilde M_0^{(\leqslant0)}\oplus\widetilde M_2^{(\leqslant1)}\oplus\widetilde M_4^{(\leqslant2)}\oplus\widetilde M_6^{(\leqslant1)}\oplus\widetilde M_8^{(\leqslant0)}\oplus\widetilde M_{10}^{(\leqslant0)}~.
\end{aligned}}
\end{equation}
Continuing to add more minimal punctures, we again see the pattern of the first few quasimodular forms having the maximum depth for a given weight, followed by depth one and zero. Therefore, for the general case with $s=0$ we propose the following general expression,
\begin{equation}
{\small
\begin{aligned}
    P_{g,0,t} \;\in\; &\widetilde M_0^{(\leqslant0)}\oplus\widetilde M_2^{(\leqslant1)}\oplus(1-\delta_{g,1}\delta_{t,1})\widetilde M_4^{(\leqslant2-\delta_{t,1})} \\[0.15cm]
    &\oplus\bigoplus_{j=3}^{g+\left\lfloor\frac{t}{2}\right\rfloor}\widetilde M_{2j}^{(\leqslant1)} \oplus\bigoplus_{k=g+1}^{3g-3+t-\left\lfloor\frac{t}{2}\right\rfloor} \widetilde M_{2k+2\left\lfloor\frac{t}{2}\right\rfloor}^{(\leqslant0)}~.
\end{aligned}}
\end{equation}
One can immediately check that this specialises to \eqref{a2_genus_nopunc} and \eqref{a2_max_min} as appropriate.

Finally, a similar result can be obtained for the cases with more maximal punctures. For a given number of maximal punctures, the Eisenstein expression generally contains the following three quasimodular form spaces with maximum allowed depth for the given weight,
\begin{equation}
    \bigoplus_{i=0}^2 \widetilde M_{6s+2i}^{(\leqslant3s+i-\delta_{t,1})}~,
\end{equation}
with some special cases accounted for by the Kronecker delta. The difference in this case compared to those with $s=0$ is that there are generally some additional quasimodular forms with smaller weights. The forms with weights less than $6s$ can themselves be organised in two groups by observing their depth after ignoring a few exceptions: (1) where the depths are given in terms of the weight $k$ as $k-3s-2+\Lambda_g+s({\rm mod}\;2)$, and (2) where the depths are given as $k-3s+\Lambda_g+s({\rm mod}\;2)$, where $\Lambda_g$ is the non-zero genus version of the $\Lambda_0$ defined above,
\begin{equation}
\Lambda_g = 2\delta_{s,0}\tilde\delta_{g,0} -
\begin{cases}
    2-s({\rm mod}\;2)&,\quad t=0~, \\[0.2cm]
    \phantom{2-}s({\rm mod}\;2)&,\quad t\neq0~.
\end{cases}
\end{equation}
This prefactor can be collected into a separate direct sum that we can generally express as
\begin{equation}\label{eq:betagst}
{\small
\begin{aligned}
    \beta_{g,s,t} &\equiv \delta_{s\%2,0}\;\tilde\delta_{s,0}\tilde\delta_{t,0}\;\widetilde M_{3s+2-\Lambda_g}^{(\leqslant0)}\oplus\tilde\delta_{s\%2,0}\tilde\delta_{s,1}\;\widetilde M_{3s+2-\Lambda_g}^{(\leqslant1)} \\[0.15cm]
    &\qquad\qquad \oplus\bigoplus_{l=1}^{\left\lfloor\frac{s}{2}\right\rfloor-1}\widetilde M_{3s+2-\Lambda_g-2\delta_{t,0}\delta_{s\%2}+2l}^{(\leqslant2l+s\%2)} \\[0.15cm]
    &\qquad\qquad \oplus\bigoplus_{m=1}^{s-1}\widetilde M_{3s+2\left\lfloor\frac{s}{2}\right\rfloor-\Lambda_g-2\delta_{t,0}\delta_{s\%2,0}+2m}^{(\leqslant2\left\lfloor\frac{s}{2}\right\rfloor-1+s\%2+2m)}~,
 \end{aligned}}
\end{equation}
where we have adopted the notation $\tilde\delta_{x,0}\colonequals1-\delta_{x,0}$ and $s\%2$ for $s({\rm mod}\;2)$.

With these preparations, at last we can present our conjecture for the direct sum of spaces of quasimodular forms in which the Eisenstein expressions of type $\mathfrak a_2$ with $g\neq0$ lie,
\begin{equation}\label{eq:big_conjecture}
{\small
\begin{aligned}
    P_{g,s,t} \;&\in\; \beta_{g,s,t}\oplus\widetilde M_{6s}^{(\leqslant3s)} \\[0.15cm]
    &\oplus(1-\delta_{g,1}\delta_{t,0})\Big[\widetilde M_{6s+2}^{(\leqslant3s+1)}
    \oplus(1-\delta_{g,1}\delta_{t,1})\widetilde M_{6s+4}^{(\leqslant3s+2-\delta_{t,1})}\Big] \\[0.15cm]
    &\quad \oplus\bigoplus_{j=3}^{g+\left\lfloor\frac{t}{2}\right\rfloor} \widetilde M_{6s+2j}^{(\leqslant3s+1)}
    \oplus\bigoplus_{k=g+1}^{3g-3+t-\left\lfloor\frac{t}{2}\right\rfloor} \widetilde M_{6s+2\left\lfloor\frac{t}{2}\right\rfloor+2k}^{(\leqslant3s)}~.
\end{aligned}}
\end{equation}
The multitude of Kronecker delta functions allow for various exceptional cases.

With this expression, for given values of $g$, $s$, and $t$, one can immediately find an Ansatz for the sum of quasimodular forms whose numerical coefficients can be fixed by looking at a finite number of terms in the $q$-series of the index. We have verified this conjecture for all examples for which there are at most nine spaces of quasimodular forms appearing in \eqref{eq:big_conjecture}.

Using the same procedure as in the $\mathfrak{a}_1$ theories, we can extract from our proposed decompositions into quasimodular forms a prediction for the order of the MLDE satisfied by Schur indices of type $\mathfrak a_2$. This general prediction is as follows,
\begin{equation}
\begin{aligned}
    {\rm Ord}^{\mathfrak a_2}_{g,s,t} &= \delta_{s+t,0}\ (2g-1) \\[0.2cm]
    &\qquad\qquad+\left(\frac{10g-6+3s+2t-\Lambda}{2}\right) \\
    &\qquad\qquad\times \left(\frac{6g-6+3s+2t+\Lambda}{2}\right)~.
\end{aligned}
\end{equation}

We have been less systematic in our analysis of $\mathfrak{a}_3$ class $\mathcal{S}$ theories, but for the examples discussed earlier we can again read off the spaces of quasimodular forms present,
\begin{equation}
{\small
\begin{aligned}
    P^{(3)}_{2,0} \;&\in\; \widetilde{\mathcal M}_0^{(\leqslant0)}\oplus\widetilde{\mathcal M}_2^{(\leqslant1)}\oplus\widetilde{\mathcal M}_4^{(\leqslant2)}\oplus\widetilde{\mathcal M}_6^{(\leqslant3)}\oplus\widetilde{\mathcal M}_8^{(\leqslant1)} \\
    &\hspace*{2.9cm}\oplus\widetilde{\mathcal M}_{10}^{(\leqslant0)}\oplus\widetilde{\mathcal M}_{12}^{(\leqslant0)} \\[0.2cm]
    P^{(3)}_{0,3} \;&\in\; \widetilde{\mathcal M}_{22}^{(\leqslant1)}(\Gamma^0(2))\oplus\widetilde{\mathcal M}_{24}^{(\leqslant3)}(\Gamma^0(2)) \\[0.2cm]
    P^{(3)}_{3,0} \;&\in\; \widetilde{\mathcal M}_0^{(\leqslant0)}\oplus\widetilde{\mathcal M}_2^{(\leqslant1)}\oplus\widetilde{\mathcal M}_6^{(\leqslant3)}\oplus\widetilde{\mathcal M}_8^{(\leqslant2)}\oplus\widetilde{\mathcal M}_{10}^{(\leqslant1)}\oplus\widetilde{\mathcal M}_{12}^{(\leqslant1)} \\
    &\hspace*{1.4cm} \oplus\widetilde{\mathcal M}_{14}^{(\leqslant0)}\oplus\widetilde{\mathcal M}_{16}^{(\leqslant0)}\oplus\cdots\oplus\widetilde{\mathcal M}_{22}^{(\leqslant0)}\oplus\widetilde{\mathcal M}_{24}^{(\leqslant0)}~.
\end{aligned}}
\end{equation}
In the first and third line these are quasimodular forms over $\Gamma_1$, and in the second over $\Gamma^0(2)$. Thus, we predict that $\mathcal I_{2,0}^{(3)}$, $\mathcal I_{0,3}^{(3)}$, and $\mathcal I_{3,0}^{(3)}$ satisfy MLDEs of order $70$, $6$, and $247$, respectively.

Finally, for the case of $\mathcal{N}=4$ super Yang-Mills indices we have the following conjecture for the general form of the Eisenstein decompositions,
\begin{equation}
{\small
    \mathcal I^{\mathcal{N}=4}_{\mathfrak a_n}(q) \in
    \begin{cases}
        \bigoplus_{i=0}^{\frac{n}{2}} \widetilde M_{2i}^{(\leqslant i)}(\Gamma_1)~, &n\equiv0\;({\rm mod}\;2)~,\\[0.3cm]
        \frac{\eta(\tau/2)^2}{\eta(\tau)^4}\bigoplus_{i=1}^{\frac{n+1}{2}} \widetilde M_{2i}^{(\leqslant i)}(\Gamma^0(2))~, &n\equiv1\;({\rm mod}\;2)~.
    \end{cases}}
\end{equation}
In this case, all the quasimodular forms that appear in the Eisenstein expression have maximum depth equal to half their weight. Once again, the order of the MLDE that these indices satisfy can be predicted \emph{\`a la} the previous discussion, and we find,
\begin{equation}
    {\rm Ord}_{\mathfrak a_n} = 
    \begin{cases}
        \frac{(n+2)^2}{4} &,\;n\equiv0\;({\rm mod}\;2)~,\\[0.2cm]
        \frac{(n+1)(n+3)}{4} &,\;n\equiv1\;({\rm mod}\;2)~,
    \end{cases}
\end{equation}
which matches and extends the results of \cite{Beem:2017ooy}.

\section{\label{sec:summary_dicussion}Further Comments}

We have seen that there is a rich interplay between Schur indices of class $\mathcal{S}$ SCFTs and the world of quasimodular and logarithmic vector-valued modular forms. This is surely only the proverbial ``tip of the iceberg'' when it comes to the modular properties of Schur indices.

Indeed, there is another vantage point on Schur index modularity that appears to be somewhat different and warrants mention. The quasimodular Eisenstein series $\mtbE_2(\tau)$ is in fact one of the simplest examples of a \emph{mock modular form} \cite{Zwegers:2008zna}. One then naturally wonders whether the world of mock modular forms and their generalisations is the more natural environment in which to situate these Schur indices. We can illustrate this for the case of Lagrangian ${\cal N}=2$ theories with $SU(2)$ gauge group and $N_f$ fundamental hypermultiplet flavours. These are Lagrangian theories so the unflavoured index for these models in presence of line defects can be written as simple contour integrals \cite{Razamat:2012uv,Gang:2012yr,Tachikawa:2015iba},
\begin{equation}
    \frac12\oint \frac{dy}{2\pi i y} \phi(y,q)_{N_f}\,\chi_\lambda(y)~,
\end{equation}
where
\begin{equation}
    \phi(y,q)_{N_f} =q^{\frac{2+N_f}{12}} \eta(\tau)^2\frac{\theta(y^2;q)\theta(y^{-2};q)}{\theta(-q^{\frac12} y;q)^{2N_f}}~,
\end{equation}
and $\chi_\lambda$ is the character of the $\lambda$-dimensional irreducible representation of $\mathfrak{su}(2)$.\footnote{We have written the expression for the index on the second sheet \cite{Cassani:2021fyv}, so after taking $q\to q e^{2\pi i}$, which is reflected in the minus sign in theta function in the denominator of $\phi(y,q)_{N_f}$.} For $N_f<4$ these theories are not conformal, but it was argued in \cite{Cordova:2016uwk,DFDZ} that nevertheless the Schur index can be meaningfully defined, and in fact related \cite{Cordova:2016uwk} to the work of Kontsevich and Soibelman on wall-crossing for BPS states \cite{Kontsevich:2008fj}. Note that $\phi(y,q)_{N_f}$ is a Jacobi form of weight one and index $4-N_f$, so the Schur index is given by combinations of Fourier modes of a Jacobi form. As such, it is expected to be a vector-valued Mock modular form \cite{Dabholkar:2012nd}, where the size of the vector is determined by the index $4-N_f$ (which is proportional to the $\beta$-function of the theory). The result is mock modular when the integrand $\phi(y,q)_{N_f}$ has poles, which happens for $N_f>1$, and otherwise it is modular.\footnote{In the latter case the index in presence of line defects forms a modular vector, which in fact was claimed to be the same modular vector as the index in presence of surface defects \cite{Cordova:2017mhb}.} For Lagrangian $\mathcal{N}=2$ theories of higher rank, we will thus expect to obtain generalisations of this mock modular structure. For example, one could perform the integrals one-by-one, starting with just the free hypermultiplet, the contribution of which is a Jacobi form. Performing a single integral we would return a form with mock modular properties, and continuing the integrations we should get higher level mock modular objects. We observe that precisely for the conformal case, the indices are quasimodular and the question is whether for the more general asymptotically free cases the mock modular perspective will be useful.\footnote{S.S.R. is grateful to A.~Dabholkar, S.~H.~Shao, and M.~Oren~Perelstein for discussions on mock modularity of Schur indices.}

We also observe that there now exist Lagrangian constructions for many strongly coupled $\mathcal{N}=2$ SCFTs, albeit ones that do not manifest all the (super)symmetries of the fixed points. This implies that many of the Schur indices studied here can be written explicitly, not just as infinite sums of the form of equations \ref{eq:a1_index_with_dedekinds} and \ref{eq:a2_index_with_dedekinds}, but also as contour integrals. The examples include, among others, many Argyres--Douglas theories \cite{Maruyoshi:2016aim} (indices given in equation \ref{eq:a1a2_AD_indices}), the $\mathfrak e_6$ SCFT \cite{Gadde:2010te,Razamat:2012uv,Gadde:2015xta,Zafrir:2019hps,Etxebarria:2021lmq} (index given in equation \ref{eq:d4e6_rank_one_indices}), and, say, the three punctured sphere of type $\mathfrak a_3$ \cite{Razamat:2019vfd} (index given in equation \ref{a3indices}). It would be very interesting to understand how quasimodularity emerges from such contour integral expressions.\footnote{One may also hope to extend the discussion more generally to ${\cal N}=1$ theories. On one hand, in some cases there exist specialisations of the $\mathcal{N}=1$ index akin to Schur limit in the sense that they reduce to integrals of theta functions \cite{Razamat:2020gcc}. On the other hand, the full ${\cal N}=1$ index is acted upon with $SL(3,{\mathbb Z})$ transformations \cite{SL3Z,Spiridonov:2012ww,Gadde:2020bov}, which suggests the possibility of embedding of all the structures discussed here into a larger and (hopefully) richer setup.}

Let us note that the Schur index has another interpretation in the framework of $2d$/$4d$ correspondences \cite{Gadde:2009kb}. Namely, it can be identified \cite{Gadde:2011ik,Alday:2013kda} with a correlation function on the corresponding UV curve of two-dimensional $q$-deformed Yang-Mills theory (in the zero area limit), which further admits a realisation in $A$-model topological string theory \cite{Aganagic:2004js}. As such, our results imply a potentially interesting modular structure to be investigated from the perspective of $q$YM and/or topological strings. 

Finally, a crucial question is precisely which subset of ${\cal N}=2$ SCFTs enjoy the simple quasimodular structure of the Schur indices investigated here. Our results could be read to suggest that this includes all untwisted class ${\cal S}$ theories (with regular punctures) in general. However, the example of Equation \ref{eq:a1a2_AD_indices} illustrates that this cannot be expected to extend to irregular theories of class $\mathcal{S}$. More generally, as these Argyres--Douglas theories can be engineered in twisted class ${\cal S}$ \cite{Beem:2020pry}, it is too much to expect the structure to carry over to twisted class $\mathcal{S}$ in general (though a generalisation of the story here could still be relevant). Similarly, we have attempted to find a relatively simple quasimodular expression for the Schur index of rank-one ${\cal N}=3$ theories \cite{Garcia-Etxebarria:2015wns,Nishinaka:2016hbw,Bonetti:2018fqz,Lemos:2016xke,Zafrir:2020epd,Agarwal:2021oyl} (which are not expected to arise in untwisted class ${\cal S}$ constructions) without success. Thus, we (at least naively) are led to believe that such models also  fall outside the set of theories in question. Giving a more precise physical account of the origin of the simple quasimodular structure observed here would be of utmost interest.

\bigskip


\begin{acknowledgments}
The authors are grateful to Yiwen Pan and Wolfger Peelaers for sharing their work \cite{PanPeelaers} and for coordinating its release with us. The work of C.B. is supported in part by ERC Consolidator Grant \#864828 ``Algebraic Foundations of Supersymmetric Quantum Field Theory'' (SCFTAlg), in part by the Simons Collaboration for the Nonperturbative Bootstrap under grant \#494786 from the Simons Foundation, and in part by the STFC consolidated grant ST/T000864/1. P.S. would like to acknowledge support from the Clarendon Fund and the Mathematical Institute, University of Oxford. The research of S.S.R. was supported in part by Israel Science Foundation under grant no. 2289/18, by a Grant No. I-1515-303./2019 from the GIF, the German-Israeli Foundation for Scientific Research and Development, by BSF grant no. 2018204, by the IBM Einstein fellowship of the Institute of Advanced Study, and by the Ambrose Monell Foundation.
\end{acknowledgments}


\appendix

\section{\label{app:modular_forms}Modular Forms and MLDEs}

In this appendix, we present important definitions and conventions for the modular objects that appear in the main text. Let $\tau$ taking values in the upper half plane $\mathfrak h$. The modular group $\Gamma_1\colonequals\mathrm{PSL}(2,\mathbb Z)$ acts on $\mathfrak h$ via M\"obius transformations,
\begin{equation}
    \gamma = \begin{pmatrix} a&b\\c&d \end{pmatrix} \in \Gamma_1~, \quad \tau \mapsto \gamma\tau = \frac{a\tau+b}{c\tau+d}~.
\end{equation}
The modular group is generated by just two elements
\begin{equation}
    T\,:\,\tau\rightarrow\tau+1~,\quad S\,:\,\tau\mapsto-\frac{1}{\tau}~,
\end{equation}
which satisfy $S^2=(ST)^3=\mathbb I$.

There are special subgroups of the modular group that have finite index and are described via congruence conditions on the matrix entries. The principal congruence subgroup of level $N$ is defined to be
\begin{equation}
    \Gamma(N) = \left\{\gamma\in\Gamma_1\;\left\vert\;\gamma\equiv\begin{pmatrix}1&0\\0&1\end{pmatrix}\mod\,N\right.\right\}~.
\end{equation}
Any subgroup of $\Gamma_1$ that includes $\Gamma(N)$ for some $N$ is called a congruence subgroup. An important congruence subgroup for our purposes is the following,
\begin{equation}
    \Gamma^0(2) = \left\{\gamma\in\Gamma_1\;\left\vert\;\gamma\equiv\begin{pmatrix}*&0\\ *&*\end{pmatrix}\mod\;2\right.\right\}~.
\end{equation}
This is generated by ${T}^2$ and ${STS}$ and is the subgroup of the modular group that leaves fixed the spin structure with anti-periodic boundary conditions on the spatial circle and periodic boundary conditions on the time circle.

A modular form, $\phi$, of weight $k$ on $\Gamma$ is a function on $\mathfrak{h}$ that is invariant under the ``weight $k$ modular action'',
\begin{equation}
    (c\tau+d)^{-k}\phi(\gamma\tau) = \phi(\tau)~,\quad \gamma\in\Gamma~.
\end{equation}
For $\Gamma=\Gamma_1$, the modular $T$ transformation implies the existence of a Fourier expansion for modular forms,
\begin{equation}
    \phi(\tau) = \sum_{n=0}^\infty a_nq^n~,\quad q\colonequals e^{2\pi i\tau}~,
\end{equation}
Whereas for $\Gamma=\Gamma^0(2)$, $T^2$ invariance gives an expansion in half-integer powers of $q$,
\begin{equation}
    \phi(\tau) = \sum_{n=0}^\infty a_nq^{\frac{n}{2}}~.
\end{equation}
The space of all modular forms of weight $k$ on $\Gamma$ is denoted by $M_k(\Gamma)$, and the ring of all modular forms over $\Gamma_1$ is freely generated over $\mathbb C$ by the weight four and weight six Eisenstein series, $\mtbE_4(\tau)$ and $\mtbE_6(\tau)$. 

More generally, the Eisenstein series $\mtbE_{2k}$ for $k\geqslant2$, are particular weight $2k$ modular forms that can be defined through their $q$ series as
\begin{equation}
    \mtbE_{2k}(\tau) = -\frac{B_{2k}}{(2k)!} + \frac{2}{(2k-1)!}\sum_{n\geqslant1}\frac{n^{2k-1}q^n}{1-q^n}~,
\end{equation}
where $B_{2k}$ is the $2k^{\rm th}$ Bernoulli number.

The ring of modular forms over $\Gamma^0(2)$ is also finitely generated by symmetric combinations of the fourth powers of Jacobi theta constants $\theta_2(q)^4$ and $\theta_3(q)^4$ over $\mathbb C$. These are defined as
\begin{equation}
    \theta_2(q) = \sum_{n=-\infty}^\infty q^{\frac12\left(n+\frac12\right)^2}~,\quad \theta_3(q) = \sum_{n=-\infty}^\infty q^{\frac{n^2}{2}}~.
\end{equation}
We can go even further to note that modular forms over $\Gamma^0(2)$ of weight $2k$ are spanned by
\begin{equation}
    \Theta_{r,s}(q) = \theta_2(q)^{4r}\theta_3(q)^{4s}+\theta_2(q)^{4s}\theta_3(q)^{4r}~,
\end{equation}
such that $r+s=k$. Alternatively, these modular forms can be described in terms of untwisted and twisted Eisenstein series, with the latter being special weight $k$ modular forms over $\Gamma^0(2)$ defined as follows,
\begin{equation}
\begin{aligned}
    {\mathbb E}_k\begin{bmatrix}-1\\ +1\end{bmatrix}(q) &= 
    -\frac{B_{k}(\frac12)}{k!}+\frac{1}{(k-1)!}\, \sum_{j=0}^{\infty} \frac{(j+\tfrac12)^{k-1}q^{j+\frac12}}{1-q^{j+\frac12}} \\
    &\quad +\frac{(-1)^k}{(k-1)!}\, \sum_{j=1}^{\infty} \frac{(j-\tfrac12)^{k-1}q^{j-\frac12}}{1-q^{j-\frac12}}~.
\end{aligned}
\end{equation}
where $B_k(x)$ is the $k^{\rm th}$ Bernoulli polynomial.

A particularly interesting and important modular form is the modular discriminant function defined as
\begin{equation}
    \Delta(\tau) = q\prod_{n=1}^\infty (1-q^n) \equiv q(q;q)_\infty^{24}~.
\end{equation}
This is a modular form of weight $12$ on $\Gamma_1$, thus expressible as
\begin{equation}
    \Delta(\tau) = 10800\Big(20\;\mtbE_4(q)^3-49\;\mtbE_6(q)^2\Big)~.
\end{equation}
The Dedekind $\eta$ function can be defined as the $24$'th root of the discriminant
\begin{equation}
    \eta(\tau) = \Delta(\tau)^{\frac{1}{24}} = q^{\frac{1}{24}}(q;q)_\infty~,
\end{equation}
which transforms as a modular form of weight $1/2$ under $\Gamma_1$ along with a phase that depends on the particular modular transformation.

The derivative of a weight $k$ modular form is no longer a modular form. However, the anomalous piece of the derivative of a weight $k$ modular form is a linear in $\tau$ and has the same form (up to a factor of $k$) as the anomalous piece of the weight two Eisenstein series,
\begin{equation}
    \mtbE_2\left(\frac{a\tau+b}{c\tau+d}\right) = (c\tau+d)^2\mtbE_2(\tau)-\frac{c(c\tau+d)}{2\pi i}~.
\end{equation}
Consequently it is possible to define a modified derivative operator---the \emph{Ramanujan--Serre derivative}---that acts on weight $k$ modular forms to return weight $k+2$ modular forms,
\begin{equation}
\begin{aligned}
    \partial_{(k)}\,:\,M_k(\Gamma) &\rightarrow M_{k+2}(\Gamma)~,\\
    f&\mapsto q(\partial_qf)+k\,\mtbE_2(\tau)f~.
\end{aligned}
\end{equation}
An $n^{th}$ order differential operator acting on a modular object of weight $k$ is thus defined by iterating the Ramanujan--Serre derivative with increasing values of $k$,
\begin{equation}
    \mathcal D^n_{(k)} \colonequals \partial_{(2n+k-2)}\partial_{(2n+k-4)}\cdots\partial_{(k+2)}\partial_{(k)}~.
\end{equation}
When we consider the action of the Ramanujan--Serre derivative on a weight zero modular function we will omit explicit subscripts.

We end by reviewing the modular linear differential equation for vector valued modular forms. A vector valued modular form $(f_1(q),\cdots,f_n(q))$ is a $n$ dimensional vector representation of the modular group by functions on the upper half space. Therefore, under a modular transformation, every element of the vector transforms into a linear combination of the others. It can be shown that every such weight $k$ vector valued modular form arises as the solution to an order-$n$ modular linear differential equation of the form
\begin{equation}
    \sum_{m=0}^{n}\left(\phi_{2(\ell+n-m)}(q)\mathcal D_{(k)}^m\right)f_i(q) = 0~.
\end{equation}
where $\phi_{2d}$ is a weight $2d$ modular form. An MLDE is said to be holomorphic or monic if the Wronksian index $\ell$ vanishes. For $\Gamma_1 $ this implies that the coefficient of the highest-order derivative term is the identity and there is no order-$(n-1)$ term due to the absence of holomorphic modular forms of weight two for $\Gamma_1$.

In the context of rational conformal field theory, this was first realised in \cite{Mathur:1988na,EguchiO:1989}. For a new mathematical review, see \cite{FrancM:2016}. The proof follows from the construction of the modular Wronskian from the vector valued modular form using the iterated Ramanujan--Serre derivatives.


\end{document}